\documentstyle[seceq,epsf,wrapft]{ptptex}

\title
{\bf Experimental study of generic billiards with microwave resonators}

\author{
Gregor {\sc Veble}$\dag$, Ulrich {\sc Kuhl}$\ddag$, Marko {\sc Robnik}$\dag$,
Hans-J\" urgen {\sc St\" ockmann}$\ddag$,
Junxian {\sc Liu}$\dag$ and Michael {\sc Barth}$\ddag$\footnote{e--mails:
gregor.veble@uni-mb.si, kuhl@physik.uni-marburg.de,
robnik@uni-mb.si,\\ stoeckmann@physik.uni-marburg.de,
junxian.liu@uni-mb.si, michael.barth@physik.uni-marburg.de}}

\inst{$\dag$ Center for Applied Mathematics and Theoretical Physics,\\
University of Maribor, Krekova 2, SI--2000 Maribor, Slovenia\\
$\ddag$ Fachbereich Physik, Universit\"at Marburg, Renthof 5,\\
D--35032 Marburg, Germany}

\abst{ In this work we study the eigenstates and the energy
spectra of a
generic billiard system with the use of microwave resonators. This is
possible
due to the exact correspondence between the Schr\" odinger equation and the
electric field equations of the lowest modes in thin microwave resonators.
We
obtain a good agreement between the numerical (exact) and experimental
eigenstates, while the short range
experimental spectral statistics show the expected
Brody-like behaviour in this energy range,
as opposed to the Berry-Robnik 
picture which
is valid only in the semiclassical region of sufficiently small effective
Planck's constant.}

\begin{document}

\maketitle

\section{Introduction}

\noindent
The subject of quantum chaos deals with the quantum mechanical properties of
classically chaotic systems, namely,
it tries to relate the classical behaviour of
Hamiltonian systems to the quantal objects such as wavefunctions and energy
spectra of their corresponding quantum mechanical counterparts.

 Due to the formal similarity of the Schr\" odinger equation to the other wave
equations of physics the concept of quantum chaos can be generalized to the
one of wave chaos. In these systems we may also observe
correspondences
between the wave mechanics and a suitably defined 'classical'
mechanics (ray dynamics),
 similar to those of the Hamiltonian systems. For example, in
the case of electromagnetic waves the appropriate 'classical' picture is the
ray optics.
Furthermore, as will be shown later, the Schr\" odinger equation of a
two-dimensional billiard system, which is a system of a free point
particle elastically bouncing off the boundaries of a given domain, and
the wave equation describing the electric field
of the low frequency modes
in a thin microwave resonator of the same planar shape are in a
one-to-one correspondence \cite{rf:Stockmann1990}.
This enables us to experimentally study the
wavefunctions and the spectra of a quantal
billiard system through the electric
field properties of its microwave counterpart. A rich variety of microwave 
billiards have been studied in the past decade. For a review and further
references see the recent book by St\" ockmann \cite{rf:Stockmann1999},
and the article by Richter 
\cite{rf:Richter1997}.

In this work we experimentally studied the quantum
properties of a family of billiards with analytical boundaries, introduced
by Robnik\cite{rf:Robnik1983}. This work is directly related and complementary
to the work done by Veble {\it et al.}
\cite{rf:Veble1999}, where the quantum mechanics of the
same system is studied numerically and theoretically. In a similar work
Rehfeld {\it et al.} studied the spectra of these billiards in superconducting
microwave cavities for different shape parameters $\lambda$ 
\cite{rf:Rehfeld1999}.
When the shape parameter $\lambda$ of the family is varied from $0$ to $0.5$,
this system undergoes a transition from a fully integrable to a fully ergodic
system.
In the transition region the system is of the mixed type, where the energy
shell of the phase space is split into regions of regular, quasiperiodic
motion, and those of chaotic motion, where almost each trajectory explores
the
whole appropriate chaotic region, in the sense that after a sufficiently
long
time it reaches any point on the chaotic component with arbitrary accuracy.
This type of motion may be called generic, as almost any randomly picked
Hamiltonian will exhibit this type of motion.

The wavefunctions and spectra of fully integrable and fully chaotic systems
show remarkable differences. In the semiclassical limit when $\hbar$ tends
to $0$ the wavefunctions of the former appear regular with their nodal lines
forming a rather
regular grid, while those of the latter appear disordered with their
amplitude being Gaussian distributed. On the other hand, the energy levels of
classically integrable systems show no correlations in this limit
 and their level spacing distribution
is Poissonian
\cite{rf:Berry1977,rf:Casati1985,rf:RobnikVeble1998},
while the energy levels of classically
chaotic systems behave statistically as eigenvalues of random matrices, with
the matrices belonging to a Gaussian orthogonal ensemble (GOE) if the system
possesses the time reversal symmetry (or some other antiunitary symmetry)
and a unitary one (GUE) in the systems
where this symmetry is broken 
\cite{rf:Bohigas1984,rf:Berry1986,rf:RobnikBerry1986,rf:Robnik1986,rf:Robnik1998}.

The semiclassical picture of a generic, mixed type system can be considered,
according to
the principle of semiclassical uniform condensation (PUSC 
\cite{rf:Robnik1998,rf:Robnik1988}), 
as composed of the independent contributions of the chaotic
and regular phase space components of its classical counterpart. In the
limit $\hbar \to 0$
almost each state can be classified as a regular or irregular, depending on
whether its Wigner function, which is the quantum analogue of the classical
phase space density, is supported by a chaotic component or an invariant
torus
in the phase space.
Although the PUSC has not yet been proven, there exists
strong numerical evidence in its support
\cite{rf:Prosen1993,rf:Prosen1994,rf:Prosen1999,rf:Prosen1995,rf:Prosen1996,rf:Prosen1998,rf:Robnik1998}.

The spectrum is then a superposition of independent
contributions stemming from
different invariant components of the classical phase
space, with the
average density of states of each contribution proportional to their
relative
classical phase space volume (Liouville measure).
The corresponding statistical properties
of each contribution are that of the GOE or GUE
for the chaotic components and Poisson for the regular ones.
This is the so called Berry-Robnik \cite{rf:Berry1984} picture.

The experiments performed in this work were, however, still far from the
semiclassical limit. Nevertheless, even in this low energy region the
wavefunctions show a certain
correspondence with the classical mechanics as they are typically
related to the shortest periodic orbits of the system. The spectral statistics
are also not expected to follow the Berry-Robnik picture in this regime.
Although lacking a deeper physical foundation, the so-called Brody
picture \cite{rf:Brody1973,rf:Brody1981}
is usually applicable to the classically
mixed type systems at these
lower energies.
It interpolates between the Poisson and the GOE (GUE)
statistics by introducing a fractional power law level repulsion between
levels \cite{rf:Prosen1993,rf:Prosen1994,rf:Robniketal1999}.
As these results have been confirmed numerically, we wanted to see
whether they can be observed in an experimental setup.

In section \ref{sec:definitions} we present the theoretical background and
some mathematical definitions. In section \ref{sec:experimental} we show the
experimental setup and explain the data interpretation. We compare the
experimentally obtained states and spectral statistics
with their numerical counterparts
in section \ref{sec:results}.

\section{Theory and Definitions}

\label{sec:definitions}
\noindent
The system we studied is a billiard \cite{rf:Robnik1983}. 
It consists of a free
particle elastically bouncing off the boundaries of a domain. In our case
the domain is two-dimensional and defined by conformally mapping the unit
circle in the complex plane $z$ by the quadratic transformation onto the
complex plane $w$ (physical plane), namely
\begin{equation}
 w(z)=z+\lambda z^2,\ w=x+i y.
 \label{eq:conformal}
\end{equation}
When $y=0$ the $x$ component ranges within the interval
$[-1+\lambda,1+\lambda]$.

The classical dynamics of the billiard systems is on the one hand very
simple
since the trajectories are just straight lines between consecutive bounces
and are independent of the energy, while on the other hand their behaviour
can
range from both integrable (such as the circular or rectangular billiard) to
completely chaotic (see e.g. Bunimovich stadium, Sinai, $\lambda=0.5$
billiards). Another type of one-parameter family of generic billiards are the 
oval billiards introduced by Benettin and Strelcyn \cite{rf:Benettin1978},
whose quantum mechanics was recently studied by Makino {\it et al.} 
\cite{rf:Makino1999}. However, this type of billiards do not have an analytic
boundary (there are points of discontinuous second derivative).

In this work we chose the parameter $\lambda=0.15$, where the system is
of a mixed type with areas of regular, quasiperiodic motion and chaotic
components sharing the phase space. The billiard shape is still convex,
so that the Lazutkin's tori and caustics still exist (the so-called
whispering gallery modes), for a review see reference~\citen{rf:Lazutkin1991}.
Therefore, the system is indeed very well described by the KAM
picture. In figure \ref{fig:SOS} we show the
surface of section (SOS) plot of the main chaotic component of this billiard,
where we plot the $x$ coordinate and the $x$ component of the momentum
unit vector (denoted by $p_x$) whenever the trajectory, which started
somewhere in the chaotic region, passes the $y=0$ line. The coordinate $x$
is taken relative to the center of the billiard at $y=0$, so that its range
is
now $x \in [-1,1]$.

The stationary Schr\" odinger equation for billiards is the Helmholtz
equation with the Dirichlet boundary conditions,
\begin{equation}
 (\Delta + k^2) \psi=0,\ \psi |_{\partial D} =0,
 \label{eq:Helmholtz}
\end{equation}
where $k^2=2mE/\hbar^2$. We are interested in the properties of the
eigenstates $\psi_i$ and the
corresponding eigenvalues $E_i$ of the
system.

To study the spectra it is convenient to unfold the energy scale so that the
density of levels on the unfolded scale is everywhere equal to one. This
removes the density of states which is an individual property of a system,
so that the possible universal properties may be observed. If the
average number of states up to a given energy $\bar{N}(E)$ is known, then
the mapping
\begin{equation}
x_i=\bar{N}(E_i)
\end{equation}
produces the unfolded spectrum with the desired (unit)
density of states. For the
2D billiard system the average number of states is given by the Weyl formula
\begin{equation}
\bar{N}(E)=\frac{1}{4\pi}\left[S k^2 -L k +K\right],
\end{equation}
where $S$ is the surface of the billiard, $L$ the length of its boundary and
$K$ corrections due to the curvature and corners of the boundary 
\cite{rf:Baltes1976}.

There are many measures of statistical properties of the unfolded spectra.
We will mainly deal with the (normalized to unity) 
distribution $P(S)$ of the level spacings
between the consecutive levels. Due to the unfolding of the spectra,
\begin{equation}
\int _0^\infty S P(S) dS=1
\end{equation}
holds. It is conjectured that, after removing the symmetries of the system,
for fully integrable systems the distribution of
levels is uncorrelated, Poissonian
\cite{rf:Berry1977,rf:Casati1985,rf:RobnikVeble1998},
leading to the distribution
\begin{equation}
P_{\rm integrable}(S)=\exp(-S),
\end{equation}
while the quantal levels of classically
fully chaotic systems are believed to behave as
eigenvalues of random matrices with matrix elements being statistically
independent and Gaussian distributed, again after the discrete geometric
symmetries of the
system are removed. If the distribution of matrix elements is
invariant under orthogonal transformations we speak of Gaussian orthogonal
ensembles (GOE) of matrices,
while Gaussian unitary ensembles (GUE) are invariant under general unitary
transformations. The GOE ensembles apply in the systems with time reversal
symmetry (such as our billiard),
and the level spacing distribution is in this case
well approximated by the Wigner distribution
\begin{equation}
 P_{\rm GOE}(S)=\frac{\pi}{2} S \exp\left(-\frac{\pi}{4} S^2 \right)
\end{equation}
exhibiting a linear repulsion between levels.

Our system is neither fully integrable nor fully chaotic. The picture of
Berry
and Robnik \cite{rf:Berry1984} states that in the semiclassical limit
$\hbar \to 0$ each of the invariant components, a chaotic component or
a regular region in phase space,
contributes an independent level sequence to the total spectrum, with the
statistical weight of individual contributions being
proportional to the phase space volume (relative Liouville measure)
of the corresponding component. The
regular components thus contribute Poissonian sequences while the
contributions
from the chaotic components are GOE(GUE)-like. When only one major chaotic
component is present (as is the case with our billiard) the distribution
$P(S)$
becomes
\begin{eqnarray}
P_{\rm BR}(S,\rho_1)=\rho_1^2\exp\left(-\rho_1 S\right)
{\rm erfc}\left(\frac{1}{2} \sqrt{\pi} \rho_2 S \right)+\nonumber\\
+\left(2 \rho_1 \rho_2 + \frac{1}{2} \pi \rho_2^3 S\right)
\exp\left(-\rho_1 S -\frac{1}{4} \pi
\rho_2^2 S^2\right),
\end{eqnarray}
where $\rho_1$ is the relative measure of the regular and $\rho_2$ of
irregular
regions in phase space, with $\rho_1+\rho_2=1$.
It goes to Poisson for $\rho_1=1$ and to Wigner for
$\rho_1=0$. It is interesting to note that $P_{\rm BR}(S=0)=1-\rho_2^2$ and
is different from zero when $S=0$ and $\rho_2<1$ ($\rho_1>0$).

However, although confirmed in the semiclassical region by numerical
computations, the
Berry--Robnik picture does not apply in the lower energy regions such as those
accessible
by our experiments for qualitatively well understood and known reasons: the
dynamical localization of chaotic eigenstates and their deviation from
uniform
extendedness.  Remarkably, a distribution introduced by Brody 
\cite{rf:Brody1973}
satisfactorily describes the behaviour of level spacings in this regime (for
not too large spacings, say $S\leq 1$)
by assuming a fractional power law level repulsion between levels,
although so far it has yet not been attributed a deeper
physical foundation. This distribution reads as
\begin{equation}
P_{\rm Brody}(S,\beta)=a S^{\beta} \exp(-b S^{\beta+1}),\ a=(\beta+1) b,\
b=\left\{\Gamma \left(\frac{\beta+2}{\beta+1}\right)\right\}^{\beta+1}
\end{equation}
and becomes Poisson for $\beta=0$ and Wigner for $\beta=1$. Unlike the
Berry-Robnik distribution, it becomes
$P_{\rm Brody}(S=0)=0$ for all $\beta\neq 0$.

This relationship of quantum eigenstates to classical mechanics can be
best studied by the use of the Wigner function 
(see reference~\citen{rf:Berry1983})
\begin{equation}
 W({\bf q},{\bf p})=\frac{1}{(2\pi \hbar)^N} \int d^N {\bf X} \exp(-i {\bf
p}
\cdot
{\bf X}/\hbar) \psi^{\dagger}({\bf q}-{\bf X}/2)\psi({\bf q}+{\bf X}/2)
\end{equation}
where $N$ is the number of degrees of freedom. It
is the quantum analogue of the classical phase space density. It is not
positive definite, however it becomes so when $\hbar\to 0$. The principle
of uniform semiclassical condensation (PUSC, 
see reference~\citen{rf:Robnik1998})
states that in this limit the Wigner function of any eigenstate
condenses uniformly onto an invariant classical object in the
phase space, which can be
either an invariant torus or a whole chaotic component. If PUSC is accepted,
the Berry-Robnik picture of level statistics becomes its direct consequence.

In our work we took the value of the Wigner function on the symmetry line
$y=0$ and integrated it over $p_y$,
\begin{equation}
\rho_{\rm SOS}(x,p_x)=\int dp_y W(x,y=0,p_x,p_y),
\end{equation}
in order to compare it with the classical SOS plot. This
yields
\begin{equation}
\rho_{\rm SOS}(x,p_x)=\frac{1}{2 \pi \hbar} \int dX \exp(-i p_x X /\hbar)
\psi^{\dagger}(x-X/2,y=0)\psi(x+X/2,y=0).
\label{eq:wsos}
\end{equation}
This 'density' is different from $0$ only for the states with even parity.
As the Wigner function and thus its projection is not positive definite but
exhibits oscillations of the order of the wavelength,
we smoothed $\rho_{\rm SOS}$ with a Gaussian, chosen narrower than the
minimum uncertainty one but still wide enough to smooth out the
oscillations.

\section{Experimental technique}

\label{sec:experimental}
\subsection{Experimental setup}
\noindent
There is an exact correspondence between the electric field of the lowest
resonances in thin microwave resonators and two-dimensional quantum
billiards.
The resonator must be shaped as a thin prism with the shape of its base the
same as that of the billiard. If we choose the $x$ and $y$ coordinates in
the plane of the base and $z$ perpendicular to it, the wave equation for
the $z$ component of electric field in the lowest $k_z=0$ TM modes is
\cite{rf:Jackson1975,rf:Stockmann1999}
\begin{equation}
 (\Delta_{(x,y)}+k^2)E_z=0,\ E_z |_{\partial D} = 0
\end{equation}
with the wavenumber $k=\frac{2\pi\nu}{c}$ ($\nu$ is the measured frequency
and $c$ the velocity of light).
The $E_z$ component of electric field clearly corresponds to the
wavefunction $\psi$ in equation (\ref{eq:Helmholtz}). By measuring the
electric field in a resonator we may then obtain the amplitude of the
corresponding wavefunction of the quantum problem
\cite{rf:Stein1992,rf:Stein1995}.

The experimental setup enabled us to measure not only the amplitude of
the wavefunction across the billiard, but also its sign. This is important
in the calculation of the experimental Wigner transforms of the
eigenfunctions - see section \ref{sec:definitions} and subsection
\ref{sec:eigenstates}. The arrangement
consisted of a lower plate into which the billiard shape was drilled with
the depth of 8mm and an upper plate. Both of the plates had a
microwave antenna inserted through them and protruding through almost the
whole depth of the billiard.
The antennae were connected
to a vector network analyzer (Wiltron 360B) capable of measuring not only the
intensity
of the signal but also its phase with respect to the input reference signal.
The upper antenna both emitted and received microwaves, whilst the lower one
was receiving them only.

The lower plate (with the drilled
billiard shape) could be moved with respect to the
upper one and thus enabled the upper antenna to reach any point in the
billiard. While only the upper antenna would be sufficient to measure the
square of the electric field in microwave resonances, the transmission
between
the antennae gives the information of the relative sign (phase)
of the electric field between them. More details of the experiment
will be described elsewhere \cite{rf:Kuhl1999}.

We prepared two resonators. The first one was drilled ($8{\rm mm}$ deep)
in a
brass 
plate and
had the shape of the whole $\lambda=0.15$
billiard,
with the scale $127{\rm mm}$ per unit of length of the billiard as defined
in equation (\ref{eq:conformal}), and was used for measuring the
wavefunctions.
The wavefunctions were measured over a quadratic grid with the separation
of points by $5{\rm mm}$.
The second resonator
was used for measuring the spectra. It represented only a half
of the billiard and thus selected only the odd parity states of the whole
billiard in order to remove the symmetry, so that the possible universal
properties of its spectrum can emerge. It was made of aluminium
with $365{\rm mm}$
per unit of length and drilled $8{\rm mm}$ deep.
When measuring the spectra we tightly screwed
the upper plate to the lower one in order to increase the conductivity
between
them and thus the quality of the resonator. Thus quality factors
of some thousand were obtained. As the determination of sign
of the electric field was not
necessary in this spectral measurement,
we used a single antenna in the upper plate, but moved it to
six different positions in order not to miss
any resonances, which can happen if the antenna lies very close to
the nodal line of a resonance.

\subsection{Interpretation of experimental data}
\label{sec:interpretation}

\noindent
The data obtained from the vector network analyzer are the complex
elements of the scattering matrix
$\cal S$ of the open channels (the two antennae in our case). The
information
we want to extract from it is the Green's function of the corresponding
quantum system,
\begin{equation}
 G({\bf r}_i,{\bf r}_j,E)=\sum_n\frac{\psi_n({\bf r}_i)\psi_n^{\dagger}
 ({\bf r}_j)}{E_n-E},
\end{equation}
which carries all the information about the spectrum and wavefunctions of
the
system. The crucial theoretical tool in analyzing our experiments is the
scattering matrix $\cal S$, which by definition connects the vector $\bf a$
of
amplitudes of incoming waves with the vector $\bf b$
of outgoing waves through the
simple relation ${\bf b}={\cal S}{\bf a}$. It was shown by Stein {\it et
al.}
 \cite{rf:Stein1995} that the scattering matrix between
the antennae at positions ${\bf r}_i$ and ${\bf r}_j$ is
related to the Green's function of the system by
\begin{equation}
 {\cal S}=(1+\alpha ^{\dagger} G)^{-1} (1+\alpha G),
\end{equation}
where $\alpha$ is a complex coupling parameter that depends on the antenna
shape and the frequency of the microwaves. As the frequency variation of the
coupling parameter is slow
it can be taken as constant close to a single resonance (within a few
widths of the resonance).

When at a given value $E$ only
a single resonance $n$ contributes to the Green's
function, the scattering matrix is simplified to
\begin{equation}
{\cal S}=1+2 \alpha \tilde{\cal G},
\end{equation}
where
\begin{equation}
\tilde{\cal G}_{ij}=\frac{\psi_n({\bf r}_i)\psi_n^{\dagger}
 ({\bf r}_j)}{E_n-E+\Delta_n-i\Gamma_n}
 \label{eq:modified}
\end{equation}
is the modified Green's function with the resonances shifted by
\begin{equation}
\Delta_n=\Im(\alpha)\sum_i |\psi_n({\bf r}_i)|^2
\label{eq:shift}
\end{equation}
in the real and
\begin{equation}
\Gamma_n=\Re(\alpha)\sum_i |\psi_n({\bf r}_i)|^2
\label{eq:width}
\end{equation}
in the imaginary direction of the energy plane.
The resonances in the scattering matrix thus appear
shifted and broadened with respect to their unperturbed positions.

The measured resonances of the $\cal S$ matrix can then be fitted with the
functions of the form $\beta/(E-\gamma)$, where $\beta$ and $\gamma$ are
complex parameters.
We should remark that a good quantitative
agreement with the numerical wavefunctions was obtained only when the full
complex information of the measured $\cal S$ matrix was used (and not only
its absolute value). Thus the full complex information is needed not
only to analyze the
transmission
data (the determination of sign) but also when fitting the complex
reflection data to obtain the wavefunction amplitude.

\section{Results}
\label{sec:results}

\subsection{Spectrum}

\noindent
The measured spectra of one half of the billiard (the odd states of the
whole billiard) were compared to the numerically calculated ones. The
difference in the level positions in the units of the mean level spacing
is shown
in the figure \ref{fig:expdiff}. The error is determined by the estimation
of the shifts $\Delta_n$ of the
resonances, by moving the receiving antenna to other positions.
In table \ref{tab:levels} we show the lowest 20 energy eigenvalues $E=k^2$ in
comparison with the 'exact' (numerical) values.

\begin{table}[t!]
\begin{center}
\begin{tabular}{|l|l|l|}
\hline
Index &Exact & Experimental \\ \hline
 1 &  13.782 &  13.805 \\
 2 &  25.208 &  25.173 \\
 3 &  38.900 &  38.902 \\
 4 &  46.265 &  46.291 \\
 5 &  55.052 &  54.981 \\
 6 &  67.785 &  67.742 \\
 7 &  73.583 &  73.518 \\
 8 &  90.826 &  90.823 \\
 9 &  94.455 &  94.335 \\
10 &  97.588 &  97.581 \\
11 & 116.537 & 116.448 \\
12 & 117.930 & 117.858 \\
13 & 129.462 & 129.402 \\
14 & 142.690 & 142.537 \\
15 & 145.705 & 145.649 \\
16 & 161.086 & 161.102 \\
17 & 168.146 & 168.151 \\
18 & 170.391 & 170.270 \\
19 & 176.678 & 176.523 \\
20 & 197.059 & 196.944 \\ \hline
\end{tabular}
\end{center}
\vspace{0.2cm}
\caption{The comparison of the lowest 20 exact and experimental
energy eigenvalues.}
\label{tab:levels}
\end{table}

The coupling gets stronger with frequency and the (absolute)
shifts increase
(see equation (\ref{eq:shift})). The frequency increase of coupling
can be shown for
the case of the ideal circular antenna \cite{rf:Stein1995}, for which
\begin{equation}
\alpha=2 \pi k R \frac{H_0^{(1)\prime}(kR)}{H_0^{(1)}(kR)},
\label{eq:alpha1}
\end{equation}
where $H_0^{(1)}$
is the zero order Hankel function.
When the diameter $R$ of the antenna is much smaller than the
wavelength (which is true in our experiments),
\begin{equation}
\alpha\approx \frac{2 \pi}{\ln (kR)}.
\label{eq:alpha2}
\end{equation}
This implies that 
the (absolute) 
difference $D_n$ between the experiment and numerics in the units of
mean level spacing  $\Delta E$ increases with the frequency.
When a large relative number of levels start to deviate by more than a few
 percent of the mean level spacing the short range level statistics such
as $P(S)$ for the actual billiard cannot be studied accurately any more, but
have merely a phenomenological value then.

Unfortunately, due to the imperfect geometry of the antennae the theoretically
expected values (\ref{eq:alpha1},\ref{eq:alpha2}) are not realized, but all 
relations remain intact if we treat $\alpha$ as a phenomenological 
parameter. This in turn implies of course that $\Delta_n$ (\ref{eq:shift}) and
$\Gamma_n$ (\ref{eq:width}) are phenomenological parameters too and therefore
the determination of the eigenenergies by fitting (\ref{eq:modified}) is 
much less accurate than in the case of ideal geometry.

The accurate range of the experimental spectra can also be
estimated independently of the numerical calculation by noting that the
shifts and widths of resonances are
similar in size
(see equations (\ref{eq:shift}) and (\ref{eq:width})). Levels can then
be considered accurate as long as the widths
of the resonances are not comparable with the mean level spacing.

The cumulative level spacing distribution
\begin{equation}
W(S)=\int_0^{S}P(S^\prime) d S^\prime
\end{equation}
for the first $180$ experimental levels is shown in the left
upper part of the
figure
\ref{fig:explevels}. In the lower left part we plot the same data in a
different representation, namely in the form of the deviations of the $U$
function \cite{rf:Prosen1993}
\begin{equation}
U(W)=\frac{2}{\pi} \arccos{\sqrt{1-W}}
\label{eq:Ufunction}
\end{equation}
from the best fitting Berry-Robnik $U$ function. This representation has the
nice and convenient
property that the statistically expected deviations of $U(W(S))$
are the same for all spacings $S$.
The experimental data clearly deviate from the best fitting
Berry-Robnik distribution and are fitted by the Brody
distribution much better, although the difference in these distributions
appears to be only slight.
The same plots for the numerically obtained
levels are shown in the right part of the figure \ref{fig:explevels}, 
giving virtually the same results.

The observed discrepancy is of course expected; the Berry-Robnik picture
applies only in the sufficiently deep (strict) semiclassical limit where
the effective Planck constant is sufficiently small. The deviation that we
observe in exact (numerical) and experimental data is attributed to the
dynamical localization of the Wigner functions of eigenstates in the phase
space, i.e. to the deviations of Wigner functions from the uniform
extendedness, as predicted by PUSC (see introduction). These effects have
been
observed and analyzed by Prosen and Robnik \cite{rf:Prosen1993}, 
further studied in references
\citen{rf:Prosen1994,rf:RobnikProsen1997,rf:Prosen1999}.
For a recent review see reference \citen{rf:Robnik1998}. 
A quantitative theory of these
(transitional) effects is not yet available.

\subsection{Eigenstates}
\label{sec:eigenstates}

\noindent
We show the comparison of a selection of
odd parity experimental states with their numerical
counterparts in figures \ref{fig:expodd} and \ref{fig:numodd}
respectively, while in figures \ref{fig:expeven} and
\ref{fig:numeven} the even parity states are shown. We plot the probability
density of
states in eight contours from $0$ to its maximum value.
The agreement between the theory is not only qualitative,
but also quantitative as individual contours of most of the plots match
nicely.
The states with the worst agreement between the theory and experiment are
those that lie in the neighbourhood of another eigenstate (within a few
typical resonance widths at the given frequency), where the perturbance of
the
antenna does not only shift the states but also mixes the
neighbouring states into linear superpositions of the unperturbed ones.
Furthermore, as the perturbance due to
the antenna is different at different points, this superposition varies at
different measuring points.

Most of the obtained states can be related to the individual shortest
periodic orbits of our system shown in figure \ref{fig:perorb}.
The two most prominent
types of states are those corresponding to
the vertical (No. 5, unstable) and horizontal (No. 1, stable)
double bounce periodic
orbit, which occur regularly in the observed range of states.
This correspondence with classical mechanics can
be best studied in the phase space.

For the states of even parity we calculated the smoothed projection
of the Wigner function as defined in (\ref{eq:wsos}). We show them in the
figure \ref{fig:wigner}, with the intersections of the relevant periodic orbit
with the SOS
shown as bullets.
The largest values of the Wigner functions do indeed cover the
periodic orbits.
Although a correspondence
to classical mechanics can be established,
the observed energy range is still far from the semiclassical limit,
where the PUSC applies. In this limit the Wigner functions are expected to
 be extended
across larger invariant objects in phase space
such as the whole chaotic components and invariant tori in the regular
regions,
and not just simple periodic orbits.
In our observed region, however, we are still in the localization regime
where
individual states are supported only by parts of these components, that is
by
the vicinity of the shortest periodic orbits of the system. This
localization
is further reflected in the spectral statistics, where the distribution of
level spacings is not the limiting Berry-Robnik but rather of the Brody
type.
The exact Wigner functions for the same states as in figure \ref{fig:wigner}
are shown in figure \ref{fig:numwigner}
using exactly the same technique, where quite
a nice agreement can be observed.

\section{Conclusion}

\noindent
The equivalence between the two-dimensional Schr\" odinger equation and the
equation for the lowest modes of thin microwave resonators enabled us to
experimentally obtain and study the spectra and wavefunctions of billiard
systems. We obtained a very good quantitative
agreement between the numerically calculated
(exact) and experimentally obtained levels and eigenstates. We must stress
again that such good agreement was only obtained by using the full complex
information of the scattering matrix between the probing antennae.

The properties of obtained spectra and eigenfunctions were in accordance with
our
expectations for a mixed type system such as ours. Although hints to
correspondence with the classical mechanics may be observed in the currently
attainable frequency range, the semiclassical limit is still out of reach of
the microwave experiments.

In conclusion, we might envision two important improvements of our experimental
approach, namely firstly, by a theoretically correct estimation of the 
complex coupling parameter $\alpha$ and secondly, by extending the scattering 
theory \cite{rf:Stein1995} of subsection \ref{sec:interpretation} to a 
two-channel theory.

\section*{Acknowledgements}

\noindent
We thank Dr. Toma\v z Prosen for assistance and advise with some computer
programs. This work was supported by the Ministry of Science and Technology
of the Republic of Slovenia and by the Rector's Fund of the
University of Maribor. It has also been strongly supported by the
scientific and academic cooperation programme between the universities of
the twin towns Maribor and Marburg.

\clearpage 
\begin{figure}[t]
\caption{The surface of section (SOS) plot for the $\lambda=0.15$ billiard
(see text for details).}
\label{fig:SOS}
\end{figure}

\begin{figure}[t]
\caption{The difference $D_n$
between the corresponding experimental and numerical
level positions of the odd parity states as a function of the
consecutive index in the units of the mean level spacing $\Delta E$, so
$D_n=(E_n^{\rm experimental}-E_n^{\rm exact})/\Delta E$.}
\label{fig:expdiff}
\end{figure}

\begin{figure}[t]
\caption{In the upper left part we show
the cumulative level spacing distribution of the first 180
experimentally obtained odd parity levels, with the best fitting Brody (dashed
line) and Berry-Robnik (thin full line)
distributions. The limiting Poisson and
Wigner distributions are shown dotted. In the lower left part we plot the
deviations of the $U(W)$ function, see eq. (\ref{eq:Ufunction}),
from the best fitting Berry-Robnik $U$ function.
The data are shown as a full curve with the thin curves representing
the statistically expected deviations, while the dashed curve is the
difference between the best fitting Brody and Berry-Robnik $U$ functions.
In the right part of the figure the same plots are shown for the corresponding
numerically obtained level range.}
\label{fig:explevels}
\end{figure}

\begin{figure}[t]
\caption{The shortest periodic orbits in the $\lambda=0.15$ billiard. The
stable orbits are represented with full lines, while unstable ones with dashed
lines.}
\label{fig:perorb}
\end{figure}

\begin{figure}[t]
\caption{A selection of the experimentally obtained odd parity states. We
draw the probability density in eight equally spaced
contours from zero to the maximum.}
\label{fig:expodd}
\end{figure}

\begin{figure}[t]
\caption{The numerical exact 
odd states corresponding to the experimental states
of figure \ref{fig:expodd}, again with eight equally spaced
probability density contours
from zero to the maximum (upper part) and the nodal lines (lower part).}
\label{fig:numodd}
\end{figure}

\begin{figure}[t]
\caption{A selection of the experimentally obtained even parity states. We
draw
the probability density in eight equally spaced
contours from zero to the maximum.}
\label{fig:expeven}
\end{figure}

\begin{figure}[t!]
\caption{The numerically obtained exact even parity states corresponding
to the experimental states of figure \ref{fig:expeven},
with eight equally spaced
probability density contours
from zero to the maximum (upper part) and the nodal lines (lower part).}
\label{fig:numeven}
\end{figure}

\begin{figure}[h!]
\caption{The smoothed projections of the experimental
Wigner functions for the even
parity states of figure \ref{fig:expeven}. We plot eight equally spaced
contours
from 0 to the maximum, with the negative value contours shown lighter. In
some plots the intersections of the relevant periodic orbit with the SOS are
shown as bullets.}
\label{fig:wigner}
\end{figure}
\newpage

\begin{figure}[h!]
\caption{The same plot as in the figure \ref{fig:wigner}, but for the
corresponding numerically obtained exact even parity states.}
\label{fig:numwigner}
\end{figure}

\end{document}